\documentclass[aps,pra,twocolumn,superscriptaddress]{revtex4}
\usepackage{bm}
\usepackage{mathrsfs}
\usepackage{amsmath}
\usepackage{amssymb}
\usepackage{graphicx}
\usepackage{amsfonts}
\usepackage{amsthm}
\usepackage{color}
\usepackage{dcolumn}
\usepackage{txfonts}

\begin{document}

\title{Fidelity Susceptibility in One-dimensional Disordered Lattice Models}
\author{Bo-Bo Wei}
\email{weibobo@cuhk.edu.cn}
\affiliation{School of Science and Engineering, The Chinese University of Hong Kong, Shenzhen 518172, China}

\begin{abstract}
We investigate quantum phase transitions in one-dimensional quantum disordered lattice models, the Anderson model and the Aubry-Andr\'{e} model, from the fidelity susceptibility approach. First, we find that the fidelity susceptibility and the generalized adiabatic susceptibility are maximum at the quantum critical points of the disordered models, through which one can locate the quantum critical point in disordered lattice models. Second, finite-size scaling analysis of the fidelity susceptibility and of the generalized adiabatic susceptibility show that the correlation length critical exponent and the dynamical critical exponent at the quantum critical point of the one-dimensional Anderson model are respectively 2/3 and 2 and of the Aubry-Andr\'{e} model are respectively 1 and 2.375. Thus the quantum phase transitions in the Anderson model and in the Aubry-Andr\'{e} model are of different universality classes. Because the fidelity susceptibility and the generalized adiabatic susceptibility are directly connected to the dynamical structure factor which are experimentally accessible in the linear response regime, the fidelity susceptibility in quantum disordered systems may be observed experimentally in near future.
\end{abstract}

\maketitle

\section{Introduction}
Quantum phase transitions (QPT)~\cite{QPT2011} occur at zero temperature when the control parameter in the Hamiltonian of a quantum many-body system is tuned to a critical value, termed quantum critical point (QCP). Quantum many-body system at its QCP exhibits scaling and universality, which states that the equilibrium properties of physical observable close to QCP can be characterized by a few critical exponents~\cite{QPT2011,Cardy1996}. To extract the entire phenomena at QCP, for instance the critical control parameter, the universal critical exponents, and the scaling functions, physical quantities borrowed from the quantum information science~\cite{QI2000}, such as the quantum entanglement~\cite{entanglementQPT2002,entanglementQPT2008} and the quantum fidelity~\cite{Zanardi2006} and the fidelity susceptibility~\cite{You2007}, have been extensively studied in various physical systems~\cite{entanglementQPT2002,entanglementQPT2008,Zanardi2006,GuReview,You2007,Zanardi2007,Venuti2007,YangMF2007,YangMF2008,Paun2008,Chen2008,Gu2008a,Gu2008b,Gu2008c,Tong2008,Gu2009,Schwandt2009,fscaling2010,fs2011,
fs2012a,fs2012b,fs2012c,fs2013a,fs2013b,fs2013c,fs2014a,fs2014b,Gu2014,fs2015a,fs2015b,You2015,Sun2017,Wei2018,Sun2018,Wang2018,You2018}. In contrast to the order parameter in characterizing phase transition, the advantage of using concepts in quantum information science in studying QPTs is that one does not need to know the microscopic symmetry of the quantum many-body systems in advance~\cite{GuReview}.

While most investigations of the fidelity susceptibility and QPTs concentrate on the traditional quantum systems driven by competing quantum Hamiltonian~\cite{Zanardi2006,GuReview,You2007,Zanardi2007,Venuti2007,YangMF2007,YangMF2008,Paun2008,Chen2008,Gu2008a,Gu2008b,Gu2008c,Tong2008,Gu2009,Schwandt2009,fscaling2010,fs2011,
fs2012a,fs2012b,fs2012c,fs2013a,fs2013b,fs2013c,fs2014a,fs2014b,Gu2014,fs2015a,fs2015b,You2015,Sun2017,Wei2018,Sun2018,Wang2018,You2018}, the fidelity and fidelity susceptibility in characterizing localization-delocalization phase transitions in quantum disordered systems~\cite{Anderson1958,AA1980,AA2009} are largely overlooked~\cite{Tong2008,spindisorder2009}. A natural question is whether the fidelity and fidelity susceptibility can be used to locate the QCP in quantum disordered systems? Whether we can extract the universal critical exponents and the universal scaling functions of the QCP in quantum disordered systems from the fidelity and fidelity susceptibility? The aim of this paper is to provide solutions to these problems.

In this work, we study the the fidelity susceptibility and the generalized adiabatic susceptibility in two paradigmatic quantum disordered lattice models, namely the 1D Anderson model~\cite{Anderson1958} and Aubry-Andr\'{e} model~\cite{AA1980,AA2009}. We show that: (i).~One can locate the quantum critical points in the 1D Anderson model and in the Aubry-Andr\'{e} model from the fidelity susceptibility and the generalized adiabatic susceptibility. (ii).~One can extract the correlation length critical exponent and the dynamical critical exponent of the QPT in the 1D Anderson model and in the Aubry-Andr\'{e} model from the finite-size scaling analysis of the fidelity susceptibility and the generalized adiabatic susceptibility. Recently two beautiful experiments~\cite{Exp2008a,Exp2008b}, one with a real-random potential (Anderson model)~\cite{Exp2008a} and the one with a quasi-periodic potential (Aubry-Andr\'{e} model)~\cite{Exp2008b}, showed that cold atoms can be employed to simulate disorder effects in quantum lattice models.  Meanwhile, the fidelity susceptibility and the generalized adiabatic susceptibility are directly connected to the dynamical structure factor~\cite{Gu2014,You2015} which are experimental accessible in the linear response regime, thus the universality of fidelity susceptibility and of the generalized adiabatic susceptibility in the disordered lattice models report in this work could be experimentally observed in cold atoms.

This paper is organized as follows. In Sec.~II, we briefly present the quantum disordered models and its quantum phase transitions. In Sec.~III, we review the physics of fidelity and fidelity susceptibility. Sec.~IV, we show the numerical results of the fidelity susceptibility in the Anderson model. In Sec.~V, we present the numerical results of the fidelity susceptibility in the Aubry-Andr\'{e} model. Finally, Sec.~VI is a discussion and summary.

\section{Quantum Disordered Lattice Models}
We consider the following disordered Hamiltonian in one-dimensional (1D) lattice
\begin{eqnarray}
H(\Delta)&=&-J\sum_{i=1}^N(c_i^{\dagger}c_{i+1}+h.c.)+\Delta\sum_{i=1}^N\epsilon_ic_i^{\dagger}c_i,
\end{eqnarray}
where $c_i$ and $c_i^{\dagger}$ are respectively the creation and annihilation operators at site $i$ with $i=1,2,\cdots,N$, $J$ is the hopping amplitude between nearest neighbor sites, $\epsilon_i$ is the onsite potential and $\Delta$ is the overall strength of the onsite potential. In the following, we take $J$ as the unity of energy. In this work, we consider two kinds of disordered models defined by different form of local onsite potential $\epsilon_i$.

The first kind of model is the Anderson like disorder~\cite{Anderson1958}, where the onsite potential distributed uniformly in the interval
\begin{eqnarray}
\epsilon_i\in[-1,1].
\end{eqnarray}
For Anderson like disorder in one-dimension, all eigenstates of the system are localized for $\Delta>0$ and all eigenstates are delocalized at $\Delta=0$. Thus there is a quantum phase transition at $\Delta_c=0$. It was shown that~\cite{Criticalexp2011} the correlation length of the 1D Anderson model at the QCP diverges as $\xi\sim |\Delta-\Delta_c|^{1/\nu}$ with the correlation length critical exponent $\nu=2/3$ and the energy gap above the ground state vanishes as $E_G\sim |\Delta-\Delta_c|^{\nu z}$ with the dynamical critical exponent $z=2$.

The second kind of disordered model is the Aubry-Andr\'{e}(AA) model~\cite{AA1980,AA2009}, where the onsite potential is quasi-periodic,
\begin{eqnarray}
\epsilon_i=\cos(2\pi\alpha i+\phi).
\end{eqnarray}
Here $\alpha=(1+\sqrt{5})/2$ is the golden ratio. The disorder like effects in the AA model come from the incommensurability between the local potential and the lattice. Aubry and Andr\'{e} shown that this model presents a QPT at $\Delta_c=2$ from a delocalized phase ($\Delta<2$) where all the eigenstates are extended to a localized phase ($\Delta>2$) where all the eigenstates are localized. For finite size lattice, it is convenient to replace $\alpha$ by $\alpha_n=F_{n+1}/F_n$ where $F_n$ and $F_{n+1}$ are two consecutive Fibonacci numbers and it is well known that $\lim_{n\rightarrow\infty}F_{n+1}/F_n=\alpha$. The lattice size can be chosen as $N=F_n$ for periodic boundary conditions. It was shown that~\cite{Criticalexp2011} the correlation length of the Aubry-Andr\'{e} model at the QCP diverges as $\xi\sim |\Delta-\Delta_c|^{1/\nu}$ with the correlation length critical exponent $\nu=1$ and the energy gap vanishes as $E_G\sim |\Delta-\Delta_c|^{\nu z}$ with the dynamical critical exponent $z\approx2.374$.

Quantum Phase transitions induced by true random disorder and induced by quasi-periodic potential are of different universality classes as demonstrated by superfluid density~\cite{Criticalexp2011}. Because the Anderson model and the Aubry and Andr\'{e} are two paradigmatic models for understanding localization transitions, this novel phase transitions well deserve theoretical studies in more physical observable. Next, we show the physics of the fidelity susceptibility and of the generalized adiabatic susceptibility and their relations to QPTs.

\begin{figure}
\begin{center}
\includegraphics[scale=0.14]{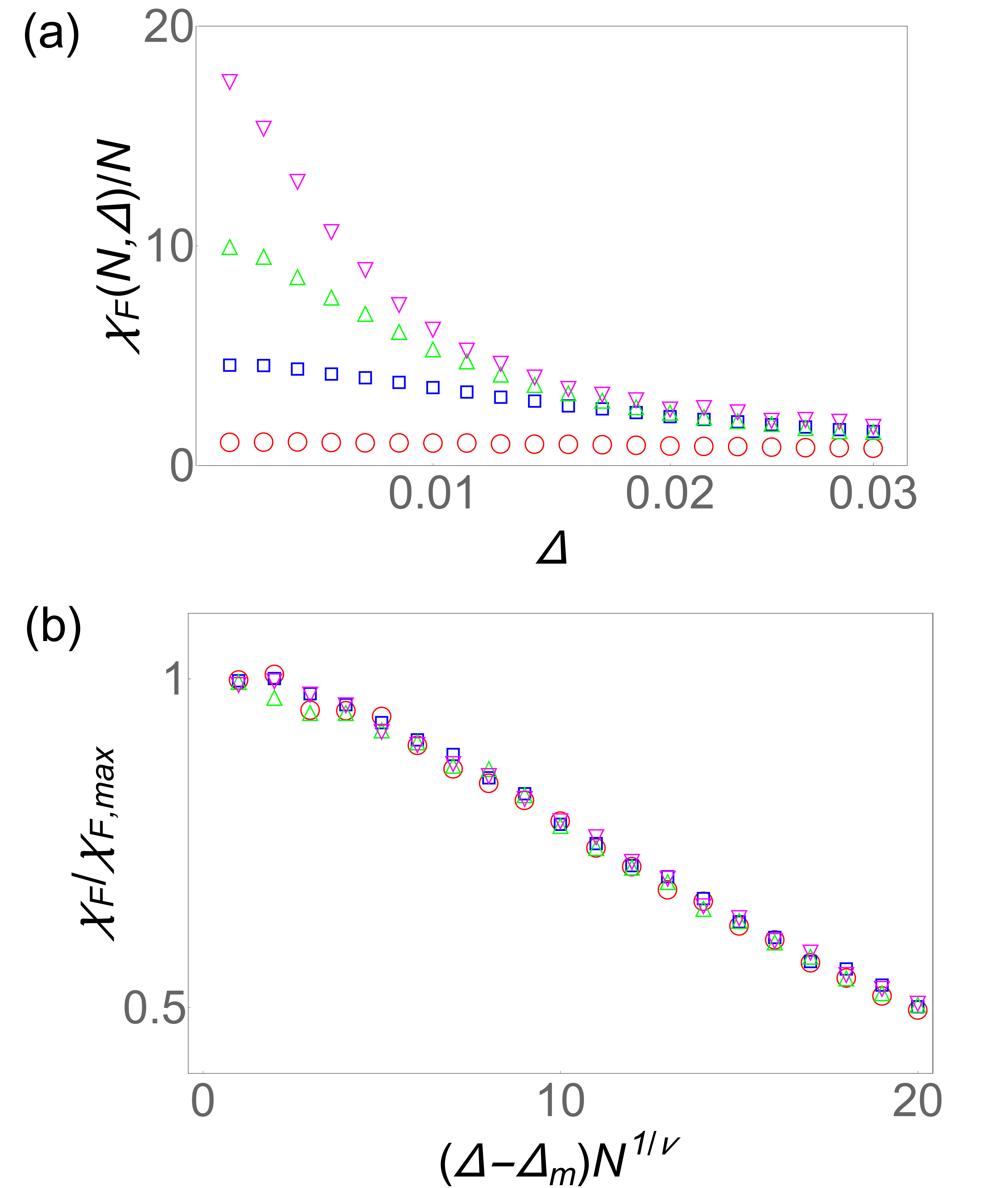}
\end{center}
\caption{(color online).~Universal finite size scaling of the fidelity susceptibility in the 1D Anderson model. (a).~Fidelity susceptibility $\chi_F(N,\Delta)$ as a function of disorder strength $\Delta$ for different lattice sizes, $N=50$ (red circle), $N=100$ (blue square), $N=150$ (green upper triangle), $N=200$ (magenta lower triangle).
(b).~Scaled fidelity susceptibility $\chi_F(N,\Delta)/\chi_{F,max}$ as a function of scaled variable $(\Delta-\Delta_m)N^{1/\nu}$. All curves for different system sizes collapse into a single curve when we choose the correlation length critical exponents $\nu=0.667$ and $\Delta_m=0$. }
\label{fig:epsart1}
\end{figure}

\begin{figure}
\begin{center}
\includegraphics[scale=0.12]{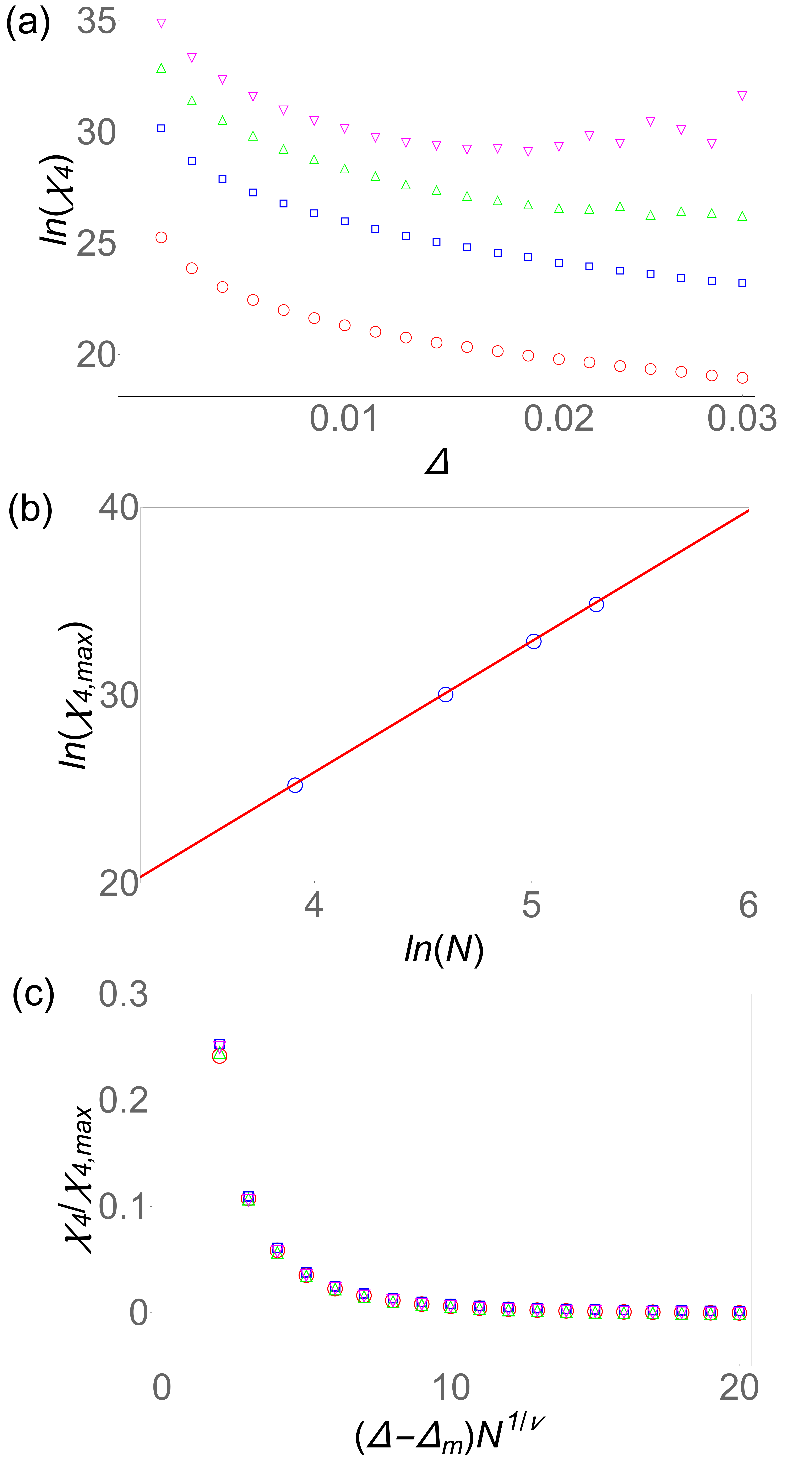}
\end{center}
\caption{(color online).~Universal finite size scaling of the generalized adiabatic susceptibility $\chi_4$ in the 1D Anderson model. (a).~The natural logarithm of the generalized adiabatic susceptibility $\chi_4(N,\Delta)$ as a function of disorder strength $\Delta$ for different lattice sizes, $N=50$ (red circle), $N=100$ (blue square), $N=150$ (green upper triangle), $N=200$ (magenta lower triangle).
(b).~The logarithm of the maximum of generalized adiabatic susceptibility as a function of the logarithm of the system sizes. Linear fit shows that $z=1.982$.  (c).~Scaled fidelity susceptibility $\chi_4(N,\Delta)/\chi_{4,max}$ as a function of scaled variable $(\Delta-\Delta_m)N^{1/\nu}$. All curves for different system sizes collapse into a single curve when we choose the correlation length critical exponents $\nu=0.667$ and $\Delta_m=0$. }
\label{fig:epsart2}
\end{figure}

\section{Fidelity Susceptibility and Quantum Phase Transitions}
Let us consider a family of many body systems with Hamiltonian
\begin{eqnarray}
H(\lambda)=H_0+\lambda H_1,
\end{eqnarray}
where $H_0$ and $H_1$ are two competing Hermitian operators and $\lambda$ is a control parameter. We assume that the many-body system described by $H(\lambda)$ undergoes a second order QPT at a critical point $\lambda=\lambda_c$. Close to the QCP, the correlation length diverges $\xi\propto|\lambda-\lambda_c|^{-\nu}$ with $\nu$ being the correlation length critical exponent and the gap above the ground state vanishes as $E_G\propto|\lambda-\lambda_c|^{\nu z}$ with $z$ being the dynamical critical exponent. The universal critical exponents $\nu$ and $z$ classify the universality of QPT and govern the universal scaling of physical observable close to QCP. In
the following, we will review how to extract the critical exponents $\nu$ and $z$ from the fidelity susceptibility and the generalized adiabatic susceptibility.

The ground state fidelity is defined as~\cite{Zanardi2006} the overlap between ground states at two different parameters $\lambda$ and $\lambda+\delta\lambda$,
\begin{eqnarray}\label{fidelity}
F(\lambda,\delta \lambda)=\left|\langle\Psi_0(\lambda)|\Psi_0(\lambda+\delta\lambda)\rangle\right|.
\end{eqnarray}
The fidelity depends on two parameters $\lambda$ and $\delta \lambda$, where $\delta\lambda$ is usually taken to be small. Because the quantum states of a many-body system within one macroscopic phase are similar, the fidelity is approximately one when two ground states are in the same phase. While the ground states at two sides of a QCP are qualitatively different, and thus one may expect that the fidelity exhibits a sharp drop at the QCP. The dominate contributions in the fidelity is the fidelity susceptibility \cite{You2007}, which may be defined as,
\begin{eqnarray}\label{fs1}
\chi_F(\lambda)=\lim_{\delta\lambda\rightarrow0}\frac{-2\ln F(\lambda,\delta\lambda)}{(\delta\lambda)^2}.
\end{eqnarray}
In the eigen states representation of the Hamiltonian, the fidelity susceptibility is \cite{You2007}
\begin{eqnarray}\label{fs2}
\chi_F(\lambda)=\sum_{n\neq0}\frac{\left|\langle\Psi_n(\lambda)|H_1|\Psi_0(\lambda)\rangle\right|^2}{[E_n(\lambda)-E_0(\lambda)]^2},
\end{eqnarray}
where $|\Psi_n(\lambda)\rangle, n=0,1,2,\cdots$ are the eigen states of $H(\lambda)$ with eigen energy $E_n(\lambda)$. Assuming that the Hamiltonian $H(\lambda)$ satisfies the eigenvalue equation, $H(\lambda)|\Psi_n(\lambda)\rangle=E_n(\lambda)|\Psi_n(\lambda)\rangle$.  Eq.~\eqref{fs1} and \eqref{fs2} can be considered as two different methods to evaluate the fidelity susceptibility.

If the control parameter is tuned as $\lambda(t)=\lambda_c+bt^r/r!\theta(t)$ with $\theta(t)$ being the step function and $b$ the adiabatic control parameter. Then the adiabatic fidelity is the overlap between the instantaneous ground state $|\Psi_0(\lambda(t))\rangle$ and the time dependent driving state $|\Psi(t)\rangle$,
\begin{eqnarray}
F(t)=\left|\langle\Psi(t)|\Psi_0(t)\rangle\right|.
\end{eqnarray}
As the energy gap at the quantum critical point vanishes, thus the system is excited by the time-dependent driving and the probability of excitations is \cite{generalized2010a,generalized2010b,noneqreview2011}
\begin{eqnarray}
P_{\text{ex}}&=&1-F(t)^2=b^2\chi_{2r+2}(\lambda_c),
\end{eqnarray}
where the adiabatic fidelity susceptibility is~\cite{generalized2010a,generalized2010b,noneqreview2011}.
\begin{eqnarray}\label{gss}
\chi_{2r+2}(\lambda)&=&\sum_{n\neq0}\frac{\left|\langle\Psi_n(\lambda)|H_1|\Psi_0(\lambda)\rangle\right|^2}{[E_n(\lambda)-E_0(\lambda)]^{2r+2}}.
\end{eqnarray}
One can see that the fidelity susceptibility is the generalized adiabatic susceptibility of order two ($r=0$). For $r=1$, we have the generalized adiabatic susceptibility of order four, $\chi_4$.

The behaviors of fidelity susceptibility at QCP have been extensively studied~\cite{Zanardi2006,GuReview,You2007,Zanardi2007,Venuti2007,YangMF2007,YangMF2008,Paun2008,Chen2008,Gu2008a,Gu2008b,Gu2008c,Tong2008,Gu2009,Schwandt2009,fscaling2010,fs2011,
fs2012a,fs2012b,fs2012c,fs2013a,fs2013b,fs2013c,fs2014a,fs2014b,Gu2014,fs2015a,fs2015b,You2015,Sun2017,Wei2018,Sun2018,Wang2018,You2018}. It was shown that the fidelity susceptibility of a finite system with size $L$ in the neighborhood of a QCP takes the universal form
\cite{Gu2008a,fscaling2010}
\begin{eqnarray}\label{fsscaling1a}
\chi_F(\lambda,L)=L^{2/\nu}\Phi_0\left((\lambda-\lambda_m)L^{1/\nu}\right),
\end{eqnarray}
where $\lambda_m$ is the control parameter at which the fidelity susceptibility is maximum, $\nu$ is the correlation length critical exponent of the QCP and $\Phi_0(x)$ is a universal scaling function as it is independent of the size of the system. From Eq.~\eqref{fsscaling1a}, the maximum of fidelity susceptibility for system with size $L$, $\chi_{F,max}\equiv max[\chi_{F}(L,\lambda)]=L^{2/\nu}\Phi_0(0)$  and thus we have
\begin{eqnarray}\label{fsscaling1}
\frac{\chi_F(\lambda,L)}{\chi_{F,max}}=\frac{\Phi_0\left((\lambda-\lambda_m)L^{1/\nu}\right)}{\Phi_0(0)}.
\end{eqnarray}
Eq.~\eqref{fsscaling1} implies that if we plot $\frac{\chi_F(\lambda,L)}{\chi_{F,max}}$ for systems of different sizes as a function of scaled parameter $(\lambda-\lambda_m)L^{1/\nu}$, then all curves of different sizes collapse into a single curve defined by $\Phi_0(x)/\Phi_0(0)$. Of course, in reality, one needs to choose $\nu$ to obtain the best data collapse. Thus fidelity susceptibility provides a simple approach to determine the universal critical exponent $\nu$ \cite{GuReview}.

The generalized adiabatic susceptibility of a finite system with size $L$ in the neighborhood of a QCP takes the universal form \cite{noneqreview2011}
\begin{eqnarray}\label{fsscaling2a}
\chi_{2r+2}(\lambda)=L^{2/\nu+2zr}\Phi_{r}\left((\lambda-\lambda_m)L^{1/\nu}\right).
\end{eqnarray}
where $\lambda_m$ is the position of the parameter at which the generalized adiabatic susceptibility is maximum, $z$ is the dynamical critical exponent, $\Phi_{r}(x)$ is a set of universal scaling functions which are independent of the size of the system. The maximum of the generalized adiabatic susceptibility for system with size $L$, $\chi_{2r+2,max}\equiv max[\chi_{2r+2}(L,\lambda)]=L^{2/\nu+2zr}\Phi_{r}(0)$ and thus we have
\begin{eqnarray}\label{fsscaling2}
\frac{\chi_{2r+2}(\lambda,L)}{\chi_{2r+2,max}}=\frac{\Phi_{r}\left((\lambda-\lambda_m)L^{1/\nu}\right)}{\Phi_{r}(0)}.
\end{eqnarray}
Eq.~\eqref{fsscaling2} tells us that if we plot $\frac{\chi_{2r+2}(\lambda,L)}{\chi_{2r+2,max}}$ for systems with different sizes $L$ as a function of scaled parameter $(\lambda-\lambda_m)L^{1/\nu}$, then all curves for different system sizes collapse into a single curve defined by $\Phi_{r}(x)/\Phi_r(0)$. In practice, one needs to choose $\nu$ to achieve the best data collapse. Thus investigations of fidelity susceptibility and of the generalized adiabatic susceptibility provide a simple approach to extracting the universal critical exponents $\nu,z$ and the universal scaling function, which determine the universality class of a QPT.

\begin{figure}
\begin{center}
\includegraphics[scale=0.115]{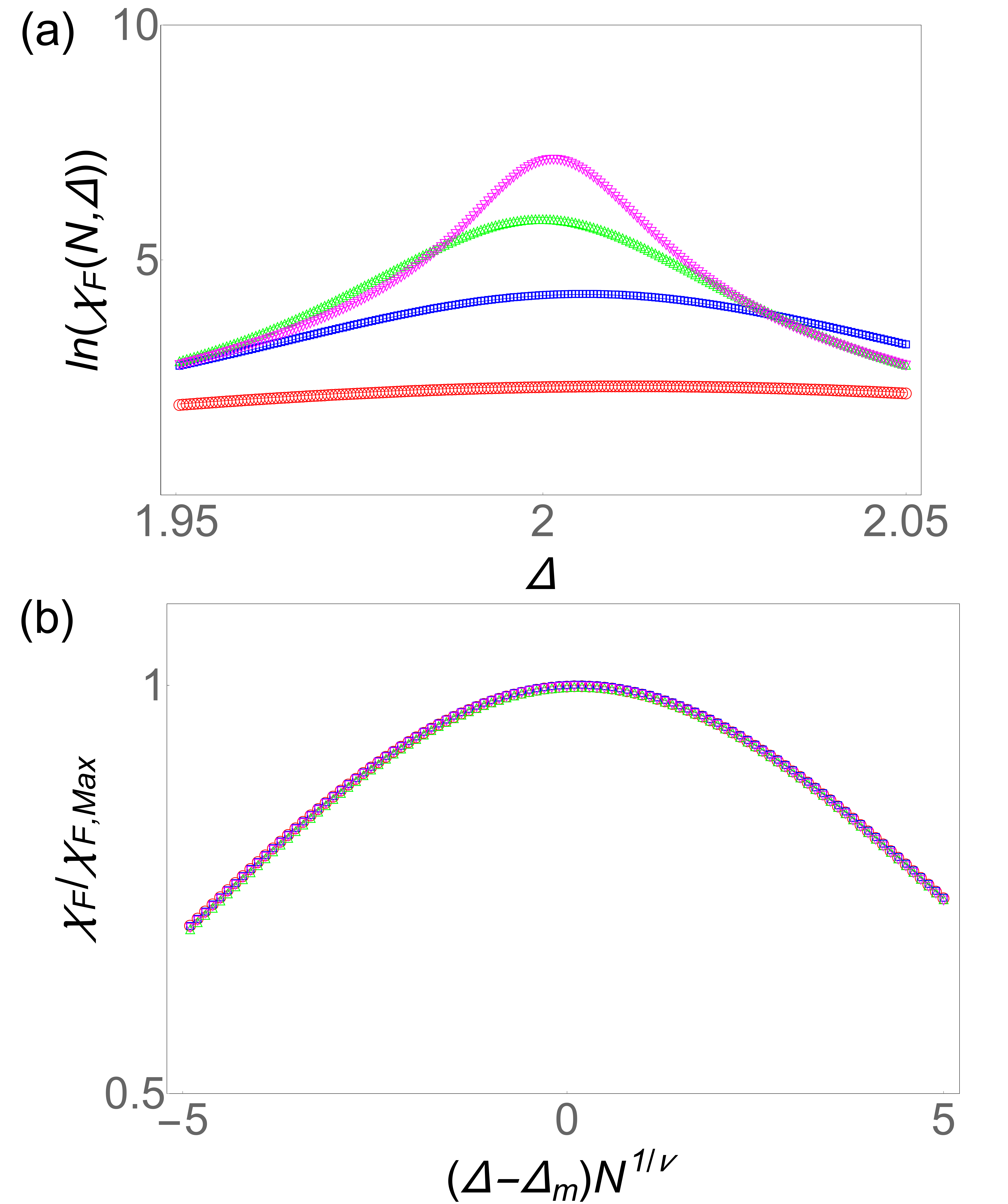}
\end{center}
\caption{(color online).~Universal finite size scaling of the fidelity susceptibility in the Aubry-Andr\'{e} model with odd number of lattice sizes. (a).~The logarithm of the fidelity susceptibility $\chi_F(N,\Delta)$ as a function of disorder strength $\Delta$ for different odd number of lattice sizes, $N=89$ (red circle), $N=233$ (blue square), $N=377$ (green upper triangle), $N=987$ (magenta lower triangle).
(b).~Scaled fidelity susceptibility $\chi_F(N,\Delta)/\chi_{F,max}$ as a function of scaled variable $(\Delta-\Delta_m)N^{1/\nu}$ with $\Delta_m$ being the position of the maximum of the fidelity susceptibility. All curves for odd number of lattice sizes collapse into a single curve when we choose the correlation length critical exponents $\nu=1.00$. }
\label{fig:epsart3}
\end{figure}

\begin{figure}
\begin{center}
\includegraphics[scale=0.14]{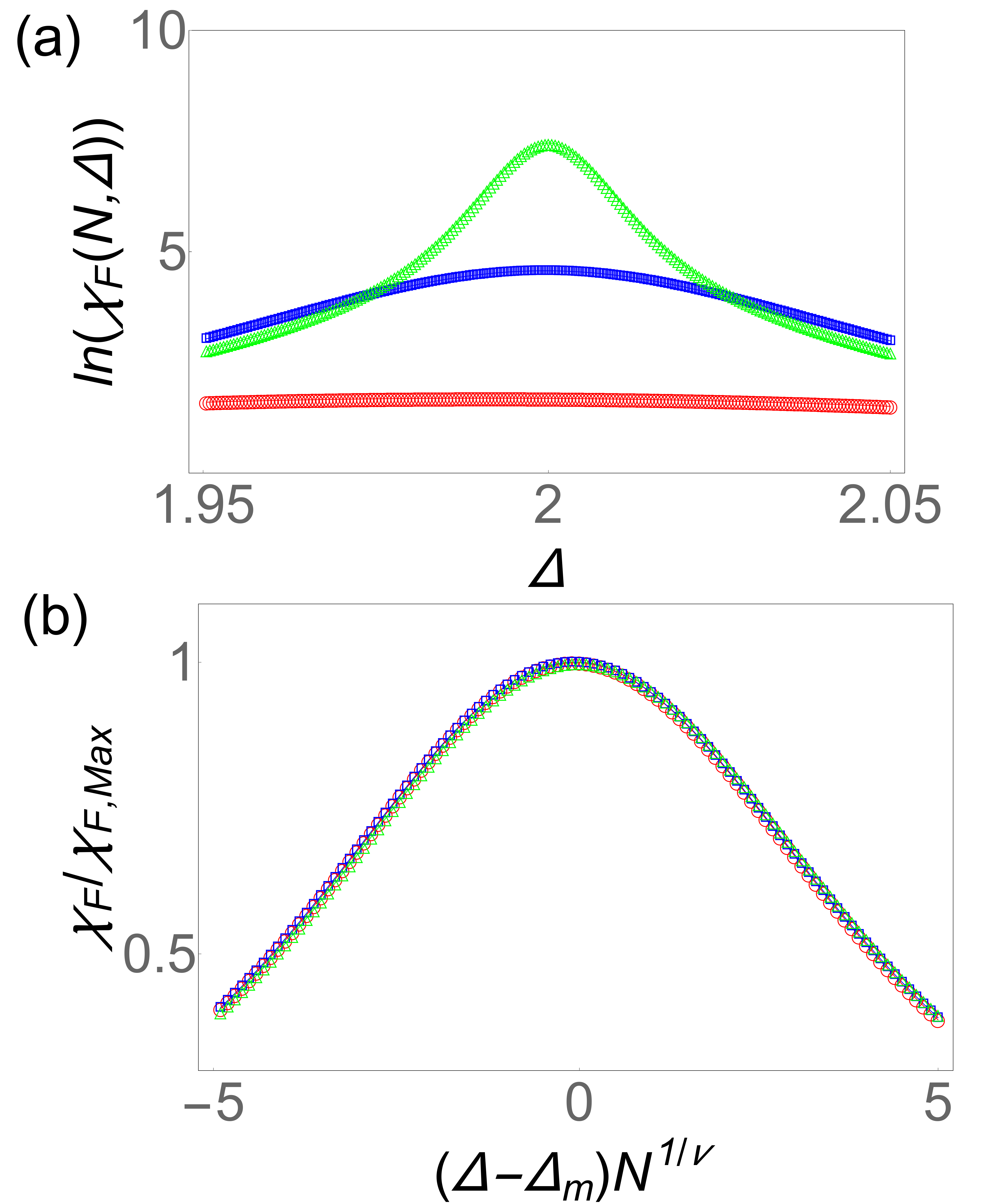}
\end{center}
\caption{(color online).~Universal finite size scaling of the fidelity susceptibility in the Aubry-Andr\'{e} model with even number of lattice sizes. (a).~The logarithm of the fidelity susceptibility $\chi_F(N,\Delta)$ as a function of disorder strength $\Delta$ for different even number of lattice sizes, $N=34$ (red circle), $N=144$ (blue square), $N=610$ (green upper triangle). (b).~Scaled fidelity susceptibility $\chi_F(N,\Delta)/\chi_{F,max}$ as a function of scaled variable $(\Delta-\Delta_m)N^{1/\nu}$ with $\Delta_m$ being the position of the maximum of the fidelity susceptibility. All curves for even number of lattice sizes collapse into a single curve when we choose the correlation length critical exponents $\nu=1.00$.}
\label{fig:epsart4}
\end{figure}

\section{The Fidelity Susceptibility in the Anderson Model}
In this section, we present the numerical results of the fidelity susceptibility in the 1D Anderson model. We calculate the fidelity susceptibility through Eq.~\eqref{fs2} and take averages over 7000 random configurations
of the local potential $\epsilon_i$. In Fig.~1(a), we show the fidelity susceptibility for 1D the Anderson model as a function of the disorder strength $\Delta$ for different lattice sizes $N=50,100,150,200$. One can see that the maximum of fidelity susceptibility always appears at $\Delta=0$, which is the QCP of the 1D Anderson model. As the disorder strength increases, the fidelity susceptibility decreases monotonically because one deviates from the QCP. Also the peak in the fidelity susceptibility increases as the size of the lattice increases. While if we plot the scaled fidelity susceptibility, namely $\chi_F(N,\Delta)/\chi_{F,max}$, where $\chi_{F,max}$ is the maximum of the fidelity susceptibility, as a function of scaled variable $(\Delta-\Delta_m)N^{1/\nu}$, then all curves for different system sizes collapse into a single one when we choose $\nu=0.667$ and $\Delta_m=0$ [Fig.~1(b)]. The extracted correlation length critical exponent is very close to the exact solution $\nu=2/3$~\cite{Criticalexp2011}.

In order to extract the dynamical critical exponent, we study the generalized adiabatic susceptibility in 1D Anderson model in Figure 2. In Fig.~2(a), we show the generalized adiabatic susceptibility $\chi_4$ as a function of disorder strength for different lattice sizes $N=50,100,150,200$. Similar to that of the fidelity susceptibility, one can see that the maximum of generalized adiabatic susceptibility also appears at the QCP $\Delta_c=0$. According to finite size scaling theory (Eq.~\eqref{fsscaling2a}), the peak in the generalized adiabatic susceptibility $\chi_{4,max}\propto N^{2/\nu+2z}$. We thus show the logarithm of the maximum of the generalized adiabatic susceptibility as a function of the logarithm of the system sizes in Fig.~2(b) and perform a linear fit to the data, i.e. $\ln\chi_{4,max}=a\ln N+b$ with $a=6.961$ and $b=1.939$. Since the correlation length critical exponent we have extracted is $\nu\approx0.667$, thus the dynamical critical exponent is $z=1.982$, which agrees to the exact solution $z=2$~\cite{Criticalexp2011}. In Fig.~2(c), we plot the scaled generalized adiabatic susceptibility, $\chi_4(N,\Delta)/\chi_{4,max}$ as a function of scaled variable $(\Delta-\Delta_m)N^{1/\nu}$, then all curves for different system sizes collapse into a single one when we choose $\nu=0.667$ and $\Delta_m=0$ [Fig.~2(c)]. Thus we get the correlation length critical exponent and the dynamical critical exponent at the QCP of the 1D Anderson model through finite size scaling analysis of the fidelity susceptibility and the generalized adiabatic susceptibility, $\nu=0.667,z=1.982$, both of them are close to the exact values~\cite{Criticalexp2011}.

\begin{figure}
\begin{center}
\includegraphics[scale=0.123]{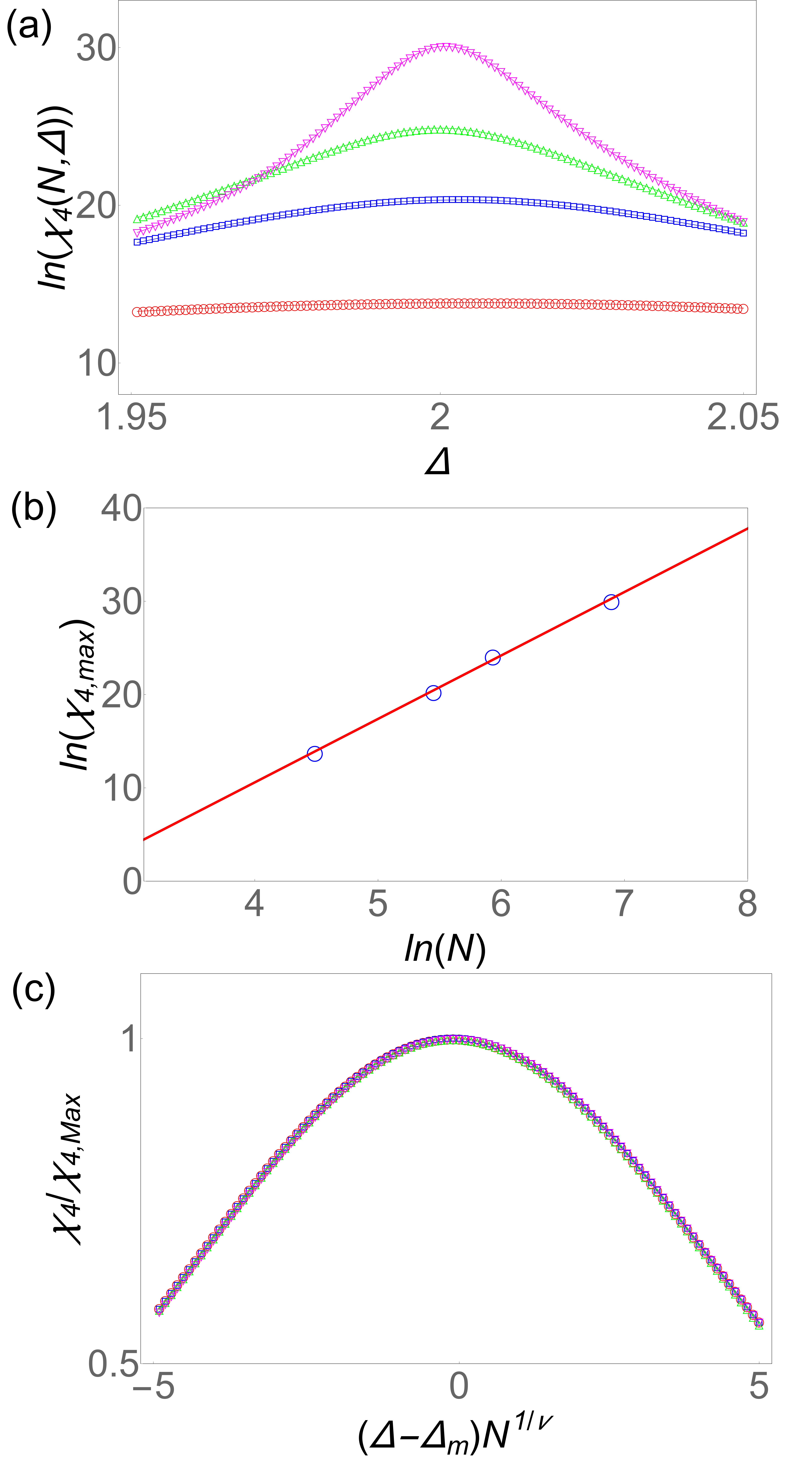}
\end{center}
\caption{(color online).~Universal finite size scaling of the generalized adiabatic susceptibility $\chi_4$ in the Aubry-Andr\'{e} model with odd number of lattice sizes. (a).~The logarithm of the generalized adiabatic susceptibility $\chi_4(N,\Delta)$ as a function of disorder strength $\Delta$ for different odd number of lattice sizes, $N=89$ (red circle), $N=233$ (blue square), $N=377$ (green upper triangle), $N=987$ (magenta lower triangle). (b).~The logarithm of the maximum of generalized adiabatic susceptibility for odd number of lattice sizes as a function of the logarithm of the system sizes. Linear fit shows that $z\approx2.38$. (c).~Scaled generalized adiabatic susceptibility $\chi_4(N,\Delta)/\chi_{4,max}$ as a function of scaled variable $(\Delta-\Delta_m)N^{1/\nu}$. All curves for odd number of lattice sizes collapse into a single curve when we choose the correlation length critical exponents $\nu=1.00$. }
\label{fig:epsart5}
\end{figure}

\begin{figure}
\begin{center}
\includegraphics[scale=0.14]{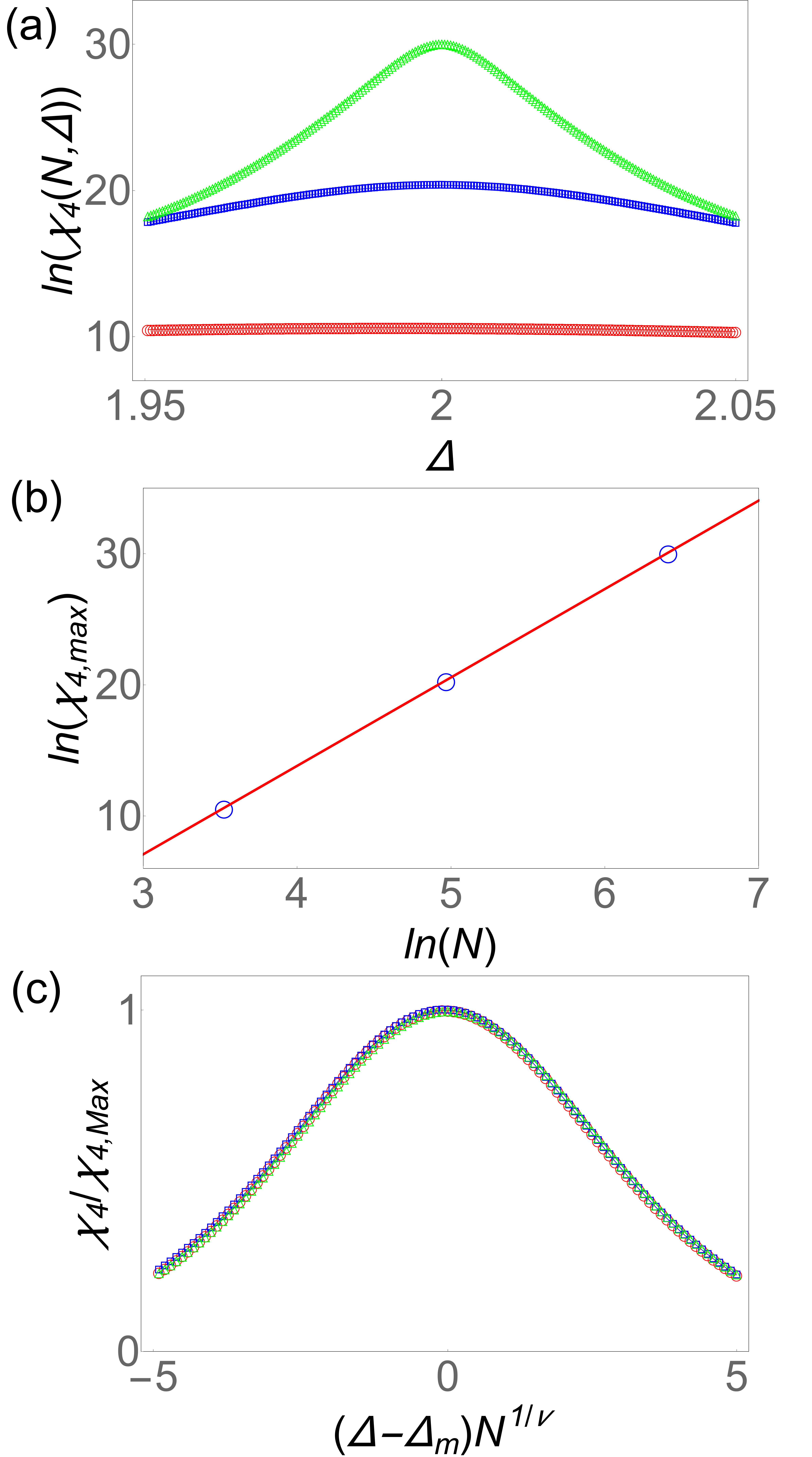}
\end{center}
\caption{(color online).~Universal finite size scaling of the generalized adiabatic susceptibility $\chi_4$ in the Aubry-Andr\'{e} model with even number of lattice sizes. (a).~The logarithm of the generalized adiabatic susceptibility $\chi_4(N,\Delta)$ as a function of disorder strength $\Delta$ for different even number of lattice sizes, $N=34$ (red circle), $N=144$ (blue square), $N=610$ (green upper triangle). (b).~The logarithm of the maximum of generalized adiabatic susceptibility for even number of lattice sizes as a function of the logarithm of the system sizes. Linear fit shows that $z\approx2.37$. (c).~Scaled generalized adiabatic susceptibility $\chi_4(N,\Delta)/\chi_{4,max}$ as a function of scaled variable $(\Delta-\Delta_m)N^{1/\nu}$. All curves for even number of lattice sizes collapse into a single curve when we choose the correlation length critical exponents $\nu=1.00$. }
\label{fig:epsart6}
\end{figure}

\section{The Fidelity Susceptibility in the Aubry-Andr\'{e} Model.}
The Aubry-Andr\'{e} (AA) model can not be analytically solved except in some extreme cases. But we can numerically exact diagonalize the Hamiltonian. Because $H(\Delta)$ is quadratic, we assume the eigenstate of $H(\Delta)$ takes the form $|\Psi\rangle=\sum_j\phi(j)c_j^{\dagger}|0\rangle$, where $\phi(j)$ is the amplitude of the wave function at site $j$. Substituting the assumed wave function into the Schr\"{o}dinger equation, we get the following system of linear equations for the amplitude of the wave function,
\begin{eqnarray}
-J\phi(j+1)-J\phi(j-1)+\Delta \phi(j)\cos(2\pi\alpha j)=E\phi(j),
\end{eqnarray}
where $j=1,2,\cdots,N$.  Diagonalizing the set of linear equations, we obtain all the eigen energies and their corresponding wave functions. Then we can numerically evaluate the fidelity susceptibility and the generalized adiabatic susceptibility in the Aubry-Andr\'{e} model through Eq.~\eqref{fs2} and Eq.~\eqref{gss}. For finite lattice sizes, one can replace $\alpha=\frac{\sqrt{5}+1}{2}$ by $\alpha_n=F_{n+1}/F_n$ where $F_n$ and $F_{n+1}$ are two consecutive Fibonacci numbers and we know that $\lim_{n\rightarrow\infty}F_{n+1}/F_n=\alpha$. Meanwhile the lattice size can be chosen as $N=F_n$ for periodic boundary conditions.

We study the finite size scaling of the fidelity susceptibility in the AA model for odd number of lattice sizes in Figure 3 and for even number of lattice sizes in Figure 4. Fig.~3(a) shows the fidelity susceptibility in the AA model as a function of the disorder strength $\Delta$ for odd number of lattice sizes $N=89$, $N=233$, $N=377$, $N=987$.  First, one can see that the fidelity susceptibility shows a peak around the QCP in the AA model at $\Delta_c=2$. Second, the peak in the fidelity susceptibility becomes more sharper as the size of the system increases. We then plot the scaled fidelity susceptibility $\chi_{F}(N,\Delta)/\chi_{F,max}$ as a function of scaled parameter $(\Delta-\Delta_m)N^{1/\nu}$ for different lattice sizes. All curves for odd number of system sizes collapse into a universal curve (Fig.~3(b)) when we choose $\nu=1.00$ and $\Delta_m$ is the parameter where the $\chi_F$ is maximum. Fig.~4(a) shows the fidelity susceptibility for even number of lattices sizes, $N=34$, $N=144$, $N=610$, as a function of control parameter $\Delta$ and Fig.~4(b) presents the scaled fidelity susceptibility as a function of scaled control parameter $(\Delta-\Delta_m)N^{1/\nu}$. To achieve data collapse, we choose $\nu=1.00$ (Fig.~4(b)). Although the universal scaling functions are different for the odd number of system sizes (Fig.~3(b)) and for even number of system sizes (Fig.~4(b)), the critical exponent for collapsing the data in two cases are the same.

To extract the dynamical critical exponent, we study the generalized adiabatic susceptibility in AA model for odd number of lattice sizes in Figure 5 and for even number of lattice sizes in Figure 6. In Fig.~5(a), we show the generalized adiabatic susceptibility $\chi_4$ as a function of disorder strength for different odd number of lattice sizes $N=89$, $N=233$, $N=377$, $N=987$. The generalized adiabatic susceptibility for even number of system sizes, $N=34$, $N=144$, $N=610$, are presented in Fig.~6(a). One can see that the maximum of generalized adiabatic susceptibility also appears at the QCP $\Delta_c=2$. According to finite size scaling theory described in Eq.~\eqref{fsscaling2a}, the peak in the generalized adiabatic susceptibility $\chi_{4,max}\propto N^{2/\nu+2z}$. We then show the logarithm of the maximum of generalized adiabatic susceptibility as a function of the logarithm of the system sizes in Fig.~5(b) and Fig.~6(b), which are respectively for odd number of lattice sizes and for even number of lattice sizes. We perform a linear fit of the data in Fig.~5(b) and Fig.~6(b),  $\ln\chi_{4,max}=a\ln N+b$ with $a=6.77, b=-16.61$ (Fig.~5(b)) and $a=6.74, b=-13.15$ (Fig.~6(b)). Because the correlation length critical exponent $\nu=1.00$, then the extracted dynamical critical exponent in the AA model is $z\approx2.38$ (odd system sizes) and $z\approx2.37$ (even system sizes). Both of them are close to the value extracted from the superfluid density $z\approx2.374$~\cite{Criticalexp2011}. In Fig.~5(c), we plot the scaled generalized adiabatic susceptibility, $\chi_4(N,\Delta)/\chi_{4,max}$ as a function of scaled variable $(\Delta-\Delta_m)N^{1/\nu}$ for all odd number of system sizes, then all curves for different odd number of system sizes collapse into a single one when we choose $\nu=1.00$. Fig.~6(c) is the same as that of Fig.~5(c) except for even number of system sizes. Thus we get the correlation length critical exponent $\nu=1.00$ and the dynamical critical exponent $z\approx2.375$ at the QCP of the AA model through finite size scaling analysis of the fidelity susceptibility and of the generalized adiabatic susceptibility.

In the AA model, the fidelity susceptibility (the generalized adiabatic susceptibility) collapse into two universal curves for odd numbers of system sizes and for even numbers of system sizes. This is of course a finite size effect. For finite lattice sizes, we replace the $\alpha$ by the ratio of two Fibonacci numbers $F_{n+1}/F_n$ and the random potential at site $i$ is $\epsilon_i=\cos(2\pi F_{n+1}/F_ni)$ for system with lattice size $N=F_n$. The profile of the random potentials are reflection symmetric about the middle site when we exclude the last site where the random potential is always fixed at one ($\epsilon_N=1$). However, for even number of lattice sizes $N=F_n$, the lattice of the system has one middle site where the strength of the random potential is always fixed at -1. While for odd number of lattice sizes $N=F_n$, there is no middle site. The two different profiles of the random potential for odd number of lattice sizes and for even number of lattice sizes govern that the physical quantities in the AA model for odd number of lattice sizes and for even number of lattice sizes collapse into two different universal functions.

\section{Summary}
In summary, we have investigated the fidelity susceptibility and the generalized adiabatic susceptibility in two paradigmatic disordered models, 1D Anderson model and the Aubry-Andr\'{e} model. Both of them present delocalization to localization quantum phase transition as the strength of the disorder increases. We found that the fidelity susceptibility is maximum close to the quantum critical point in both models, through which one can locate the quantum critical point in disordered systems. Finite-size scaling analysis of the fidelity susceptibility and of the generalized adiabatic susceptibility show that the correlation length critical exponent and the dynamical critical exponent at the QCP of the 1D Anderson model are $\nu=2/3$ and $z=2$, which are respectively $\nu=1$ and $z=2.375$ in Aubry-Andr\'{e} model. The fidelity susceptibility and the generalized adiabatic susceptibility are directly connected to the dynamical structure factor~\cite{Gu2014,You2015} which are experimentally accessible through linear response theory. Recently two experiments~\cite{Exp2008a,Exp2008b}, one with a real-random potential (Anderson model)~\cite{Exp2008a} and the other with a quasi-periodic potential (Aubry-Andr\'{e} model)~\cite{Exp2008b}, showed that the disorder effects in quantum lattice models can be simulated in cold atoms, the universality of fidelity susceptibility in quantum disordered systems may be observed experimentally in near future. Besides, we have investigated the quantum phase transitions induced by disorder in non-interacting systems from the fidelity susceptibility approach and it would be very interesting to investigate the fidelity and the fidelity susceptibility in the many-body localization transitions~\cite{MBL1,MBL2,MBL3,MBL4,MBL5,MBL6,MBL7} and the fidelity susceptibility approach may be able to extract accurate universal critical exponents at the many-body localization transitions and clarify some unsolved issues there~\cite{MBL7}.

\begin{acknowledgements}
This work was supported by the National Natural Science Foundation of China (Grant Number 11604220) and the President's Fund of the Chinese University of Hong Kong, Shenzhen.
\end{acknowledgements}

\end{document}